\documentclass{article}
\usepackage{PRIMEarxiv}
\usepackage[utf8]{inputenc} 
\usepackage[T1]{fontenc}    
\usepackage{hyperref}       
\usepackage{url}            
\usepackage{booktabs}       
\usepackage{amsfonts}       
\usepackage{nicefrac}       
\usepackage{microtype}      
\usepackage{lipsum}
\usepackage{fancyhdr}       
\usepackage{graphicx}       
\graphicspath{{media/}}     
\usepackage{natbib}
\usepackage{amsfonts,amsmath,amssymb}
\usepackage[table]{xcolor}

\usepackage{subfigure,bm,yfonts,cleveref,xcolor}

\title{Integrating U-nets into a Multi-scale Waveform Inversion for Salt Body Building}

\author{
  Abdullah Alali \\
  KAUST \\
  Saudi Arabia, Thuwal \\
  \texttt{abdullah.alali.1@kaust.edu.sa} \\
  \And
  Tariq Alkhalifah \\
  KAUST \\
  Saudi Arabia, Thuwal \\
  \texttt{tariq.alkhalifah@kaust.edu.sa} \\
}

\newcommand{\norm}[1]{\left\lVert#1\right\rVert}
\usepackage{subfigure,bm,yfonts,cleveref,xcolor}

\newsavebox{\measurebox}

\begin{document}
\maketitle


\begin{abstract}
Full-waveform inversion (FWI) has the potential to provide high-resolution models of the Earth, even in complex media. However, it  is most likely to fail when starting with a poor initial model, especially if the data lack low frequencies and long offsets. For salt provinces, these issues are heightened when the initial model lacks any prior information of the salt. Conventionally, salt bodies are often included in the FWI starting model by delineating the top and the base of the salt performing multiple tomography and imaging applications commonly known as the top-down workflow. However, the top-down workflow can be time-consuming and depends on human interpretation, which is prone to error. Studies show that FWI can improve the result obtained by the top-down workflow provided that the data are recorded using long offsets, and contain low frequencies, which is not always available. Thus, we develop an approach to invert for the salt body starting from a poor initial model, limited data offsets, and the absence of low frequencies. We leverage deep learning, three U-net networks, to apply multi-stage flooding and unflooding of the velocity model. We take advantage of the multi-scale frequency scheme of FWI and apply the networks on the inversion result after each frequency scale. Specifically, starting from an initial velocity, we apply a multi-scale FWI using three frequency bandwidths from low to high frequency. The networks after the first two bandwidths are trained to flood the salt and the network after the last frequency bandwidth is trained to unflood it. Since flooding and unflooding are inherently a 1D process and to reduce cost in generating training data and in the training itself, we use 1D U-net for the three networks. We create abundant 1D velocity models with and without salt layers and perform FWI using poor initial models to generate the training datasets. We verify the method on the synthetic BP 2004 salt model benchmark. We only use the synthetic data of short offsets up to 6 km and remove frequencies below 3 Hz. We show that the method can correctly recover the salt model starting from a poor model even with the data offset and frequency limitations. We also apply the method to real vintage data acquired in the Gulf of Mexico region. The real data lack frequencies below 6 Hz and the streamer length is only 4.8 km. With these limitations, we manage to recover the salt body and verify the result by using them to image the data and analyze the resulting angle gathers.
\end{abstract}

\section*{Introduction}
Salt bodies are massive geological structures with irregular shapes. They are known to form potential hydrocarbon traps. Moreover, in recent years, salt bodies have attracted interest in storing hydrogen H$_2$ as a green source of energy \citep{muhammed2022saltcavern}. Therefore, the energy industry is keen to obtain high-quality seismic images for salt models. Successful imaging for salt structures requires an accurate velocity model, which is hard to obtain for various reasons. They are often associated with strong multiples, especially for shallow salt bodies. They possess distinct physical properties compared to the surrounding sediments resulting in high acoustic impedance reflecting most of the seismic energy. The steeply dipping complex nature of the salt bodies raises the need to use advanced algorithms, such as those based on solving of the wave equation. 
\\
\\
Full-waveform inversion (FWI) is a non-linear optimization problem that reconstructs a high-resolution velocity model for the subsurface \citep{tarantola1984inversion}. It is implemented by minimizing the difference between the recorded seismic data and modeled synthetic data. Due to the sinusoidal nature of waves, FWI is likely to fall into a local minimum when the recorded data are far from the synthetic ones by more than half a cycle, an issue known as the cycle skipping problem \cite[e.g.,][]{chi2014full,guo2017velocity,hu2018retrieving}. This often happens when the starting model is far from the true one. The cycle skipping is more severe for high frequencies due to the high oscillatory nature of seismic data. To mitigate the cycle-skipping, it is common to perform the inversion using a multi-scale strategy, where the low frequencies are used first in the inversion, and the higher ones are included progressively \citep{bunks1995multiscale}. 
\\
\\
FWI shows high potential in reconstructing salt bodies in various fields such as in the Gulf of Mexico (GOM) \citep{shen2017Atlantis,wang2019saltComingAge}, in Brazil \citep{gans2021brazilsalt}, and in West of Africa \citep{gabrielli2021gabon}. The success of FWI in salt regions relies on two main procedures. The first procedure involves starting from a model containing an initial salt geometry obtained by the known top-down workflow \citep[e.g,][]{shen2017Atlantis,zhang2018correcting}. The top-down workflow is time-consuming and prone to human error. It consists of the following steps: applying traveltime tomography or FWI to recover the pre-salt sediments, interpreting the top of the salt (\textbf{ToS}) from a seismic image, flooding the salt velocity in depth from the picked salt, and delineating the salt base from a seismic image \citep{dellinger2017garden}. In complex salt bodies, multiple iterations of the top-down workflow are needed and different scenarios are proposed. The second procedure involves using advanced acquisition technology such as ocean bottom node (OBN) that enables long offsets, full azimuth, and low frequencies. The significance of low frequencies was first demonstrated by \cite{brenders2007waveform} as they inverted for the BP 2004 synthetic salt model using frequencies down to 0.5 Hz. However, seismic data often do not record usable energy below 2 Hz. In a real data application, \cite{shen2017Atlantis} and \cite{michell2017automatic} applied FWI in the Atlantis field with frequencies starting from 1.6 Hz and offsets exceeding 20 km. They managed to correct for misinterpreted salt obtained by the top-down approach. Similarly, \cite{wang2019saltComingAge} reshaped the salt model of the Keathly Canyon using FWI with minimum frequency and maximum offset of 2.25 Hz and 18 Km, respectively. However, we are often bounded by the cost that limits the acquisition aperture, and by the physical nature of waves and surrounding noise that limit the amount of usable low frequencies. 
\\
\\
Recently, deep learning has been utilized in many geophysical applications including modelling, processing, interpretation, and inversion, overcoming many limitations of the conventional methods \citep{huang2022high,song2021solving,liu2022coherent,harsuko2022application,alali2022seismic,zhou2020salt,xiong2018seismic,kazei2020deep,yang2019deep,araya2018deep}. In salt model building, deep learning contributes through two general approaches. The first approach automates the interpretation in the top-down workflow. \cite{gramstad2018automatedTopBase} suggested training two individual convolutional networks (CNNs) to pick the top and the base of the salt. \cite{sen2020saltnet} shows that CNNs can detect a good initial salt model from noisy low-resolution seismic images. \cite{zeng2019automatic_salt}, \cite{zhou2020salt}, and \cite{naeini2020deep-salt-unet} 
 proposed to interpret salt bodies using U-net, an encoder-decoder CNN architecture, which has become a common architecture in many interpretation tasks. Due to the erroneous velocity at the beginning of the velocity model building, the reflectors are likely to appear in innacurate positions within the image, which leads to inaccurate detection for salt bodies. Thus, \cite{zhao2021automatic-RTM-FWI-salt} suggested an iterative salt picking from reverse time migrated (RTM) images followed by FWI. Generally, this class of methods requires advanced imaging, which is often applied using high frequencies, fine grids, and higher wavefield solution costs. The second approach focuses on improving the inversion algorithm such that it recovers the salt without cycle skipping. \cite{asnaashari2013regularized} regularized the inversion by adding prior salt information obtained by training a network on seismic images. \citep{yang2019deep} and \cite{araya2018deep} proposed a direct mapping by a network from shot gathers to velocity models containing salt bodies. As a result of performing supervised training, their approach did not provide accurate results when applied on different datasets. 
 \\
 \\
In this paper, we aim to achieve three goals: automate the picking of salt top/base for the flooding and unflooding processes, reduce the cost of salt model building by eliminating the high-frequency imaging cost, and mitigate the low frequency and large offset requirements in FWI. We propose to apply a multi-scale FWI starting from a poor model such as a constant velocity or a linearly increasing model with depth. After each bandwidth scale we apply a network for flooding or unflooding. We somewhat combine the two approaches mentioned above as we adapt the top-down approach but within the inversion algorithm the need for imaging. We train two networks for flooding, which are applied in the first two bandwidth scales, and one for unflooding used in the last scale. Our work complements \cite{alali-unflooding}, where they focus only on the unflooding and subsalt inversion. We adapt a 1D U-net architecture and train the network using 1D models as the flooding and unflooding are naturally a 1D process. 
\\
\\
This paper is structured as follows. We start by giving an overview of the theory of FWI. We then introduce our proposed workflow and explain the network setup and the generation of the training data. We then provide two applications: a 2D synthetic example and a 2D vintage real data, and conclude by discussing the results.

\section*{Full-wavefrom Inversion}
Full-waveform inversion (FWI) is an ill-posed non-linear optimization problem that recovers the unknown subsurface properties \citep{tarantola1984inversion,virieux2009overview}. Starting from an initial model, FWI updates the model ($\textbf{m}$) iteratively by minimizing the least squared difference between observed seismic data ($\mathbf{d^{obs}}$), and the synthetic ones. To alleviate the ill-posedness of FWI, prior geological knowledge is often imposed by preconditioning, constraints, and regularization techniques \citep{xue2017full,kazei2017salt,esser2016constrained}. The objective function for FWI can be formulated as: 
\begin{equation}
    J_{FWI}(\mathbf{\textbf{m}}) = \norm{\mathbf{d^{obs} - \mathcal{F}(\textbf{m})}}^2_2 + \lambda \mathcal{R}(\textbf{m}),
    \label{eq:fwi}
\end{equation}
where $\mathcal{R}$ is a regularization term and $\lambda$ is a regularization coefficient. Here, $\mathcal{F}$ is the modeling operator for an acoustic constant-density wave equation given by:
\begin{equation}
    (\nabla^2 - \frac{1}{\textbf{m}^2} \frac{d^2}{dt^2}) \hspace{1mm} \textbf{u}=\textbf{f},
    \label{eq:we}
\end{equation}
where the synthetic data are the projection of the wavefield ($\textbf{u}$) on the surface and $\textbf{f}$ is a source function.
\\
\\
In salt body FWI, a common choice for $\mathcal{R}$ in equation~\ref{eq:fwi} is the total variation regularization (TV) \citep{kalita2019regularized,kazei2017salt}, which is also used as constraints in some other implementations \citep{esser2018total,esser2016constrained}. TV aims to minimize the difference between two adjacent points, which introduces smoothness in the model while keeping sharp boundaries such as those in salt bodies. Assuming a 2D case, the TV norm is given by: 
\begin{equation}
    \norm{\mathbf{m}}_{TV} = \norm{\nabla \mathbf{m}} = \frac{1}{h} \sum_{x,z} \sqrt{ (m_{x+1,z}-m_{x,z})^2 + (m_{x,z+1}-m_{x,z})^2 + \epsilon},
    \label{eq:TV}
\end{equation}
where ($x$) and ($z$) represent the grid points in the horizontal and depth directions, respectively, and ($h$) is the grid spacing. TV regularization is not differentiable at zero variation. Therefore, a small number ($\epsilon$) is added for stability. 
\\
\\
FWI is commonly solved by gradient methods, which are based on linearization of the optimization problem. Linearizing the wave equation admits a single scattering solution, which is known as the Born approximation. As a result, FWI often fails to converge to the correct model when the initial model is far from the ground truth. We can adhere to the Born assumption by using a multi-scale strategy, where the inversion is performed using low frequency bandwidths and progressively including the higher ones \citep{bunks1995multiscale}. The relationship between the frequencies in the data and the model update is well disseminated \citep{alkhalifah2016full}. When low enough frequencies are present in the data, the large structures corresponding to low wave-numbers can be reconstructed enhancing the initial model for the next bandwidth of frequencies. In addition, recording diving waves in the data with long enough offsets contributes largely in updating the low wave-number components \citep{vigh2013long,sirgue2006importancelong}. In salt regions, low frequencies and long offsets are crucial to invert for the salt bodies, especially when we do not include any salt information in the initial model. Unfortunately, in many cases, low frequencies and long offsets are absent from the data, which underlined the necessity for what we will propose here

\section*{Method}
Here, we present an approach to invert for the salt model starting from a poor initial model that lacks salt information and using data recorded with short offsets and no low frequencies. With these conditions, applying equation~\ref{eq:fwi} will fail to reconstruct the salt body. Instead, it will reveal energy indicating the top of the salt, like in imaging. In fact, it will behave like least squares imaging applied with a generally inaccurate velocity. Thus, due to the erroneous starting model, the recovered top of the salt is likely to be incorrect. However, we can feed the inverted model to a network trained to correct for the pre-salt velocity and flood the salt. When flooding is properly excuted in the model, applying FWI will focus the energy at the base of the salt. Therefore, we can similarly train a network to unflood the salt from the base and estimate the sub-salt velocity.    
\\
\\
\subsection*{The neural network setup}
\begin{figure}[!htb]
    \centering
    \includegraphics[width=0.4\textwidth]{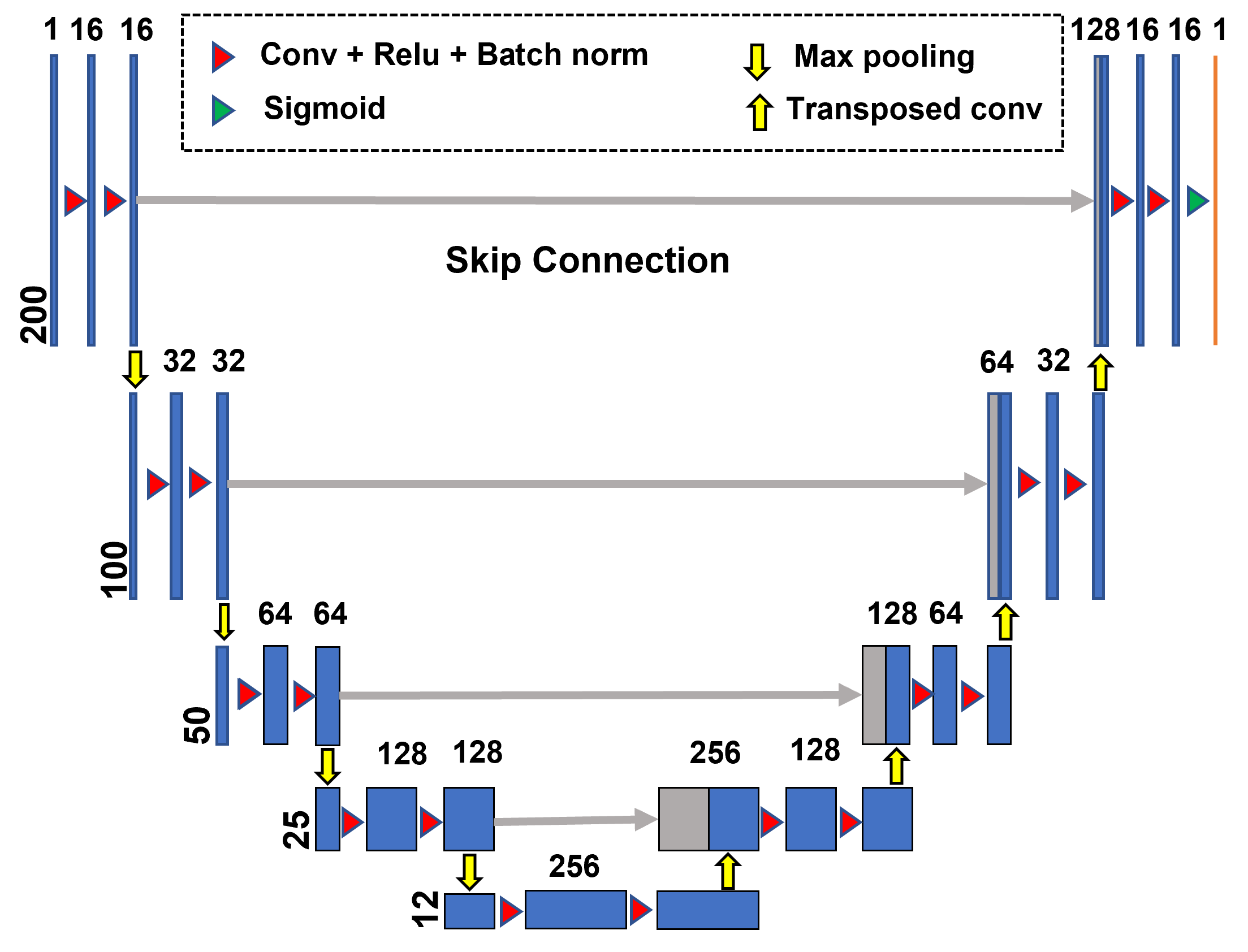}
    \caption{The One-dimensional U-net architecture.}
    \label{fig:unet}
\end{figure}

U-net is a neural network architecture that consists of an encoder and decoder convolutional blocks. It was originally developed in the medical imaging community \citep{ronneberger2015unet} for biomedical image segmentation. Then, it was adapted to many geophysical tasks such as in interpretation \citep{naeini2020deep-salt-unet}, denoising \citep{birnie2021self} and inversion \citep{alali-unflooding}. The most prominent attraction of U-net is that it extracts features from the input in different compressed levels in the encoder block. Then, it returns the compressed input to its original size in the decoder block. 
\\
\\
Here, we adopt a one-dimensional (1D) U-net architecture \citep{li2022high,alali-unflooding} to invert for the salt bodies. The architecture is displayed in Figure~\ref{fig:unet}. It is composed of four levels of encoding-decoding. Each level of the encoder is composed of two CNN layers combined with rectified linear unit (Relu) activation function and a batch normalization operator. A max-pooling operator between different encoder levels is used to reduce the dimensions by a factor of 2. Similar components make up the decoder part except for a transposed CNN layer to recover the input size instead of the max-pooling. A sigmoid function is applied after the final layer to produce the output. 
\subsection*{Training workflow}
\begin{figure}[!h]
    \centering
    \includegraphics[width=0.9\columnwidth]{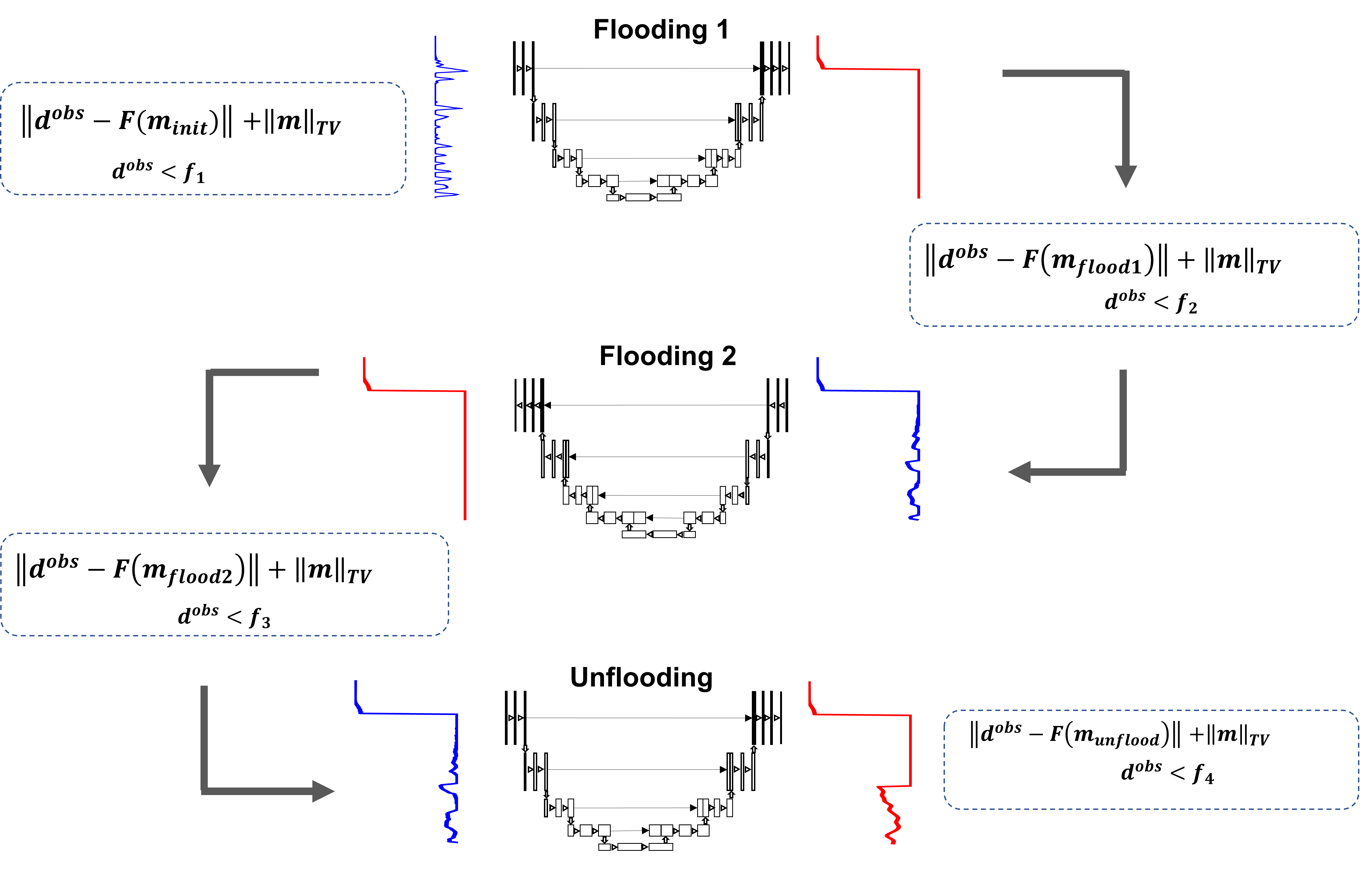}
    \caption{The workflow of the multi-scale FWI with U-net for salt reconstruction.}
    \label{fig:workflow}
\end{figure}

We propose to train three U-nets: two for flooding and one for unflooding (i.e., U-net1: flooding, U-net2: flooding, U-net3: unflooding) to aid FWI in building the salt model. Figure~\ref{fig:workflow} shows a demonstration of the approach. We apply FWI in a multi-scale fashion starting from a poor initial model ($\mathbf{m_{init}}$). We utilize the networks as a post-inversion correction after each frequency bandwidth. We apply a low-pass filter to the data using three cut-off frequencies: $f_1, f_2, f_3$ such that $f_1< f_2< f_3$. U-net1 takes the result of the first inversion bandwidth ($d^{obs}$ frequency $< f_1$), partly corrects the post-salt velocity and floods the salt. The subsequent inversions with the higher bandwidth ($d^{obs}$ frequency $< f_2$) should improve the salt's top dramatically as it starts with a much better model. Then, U-net2 is used to apply flooding again as the overall model above the Salt becomes more mature. After the third bandwidth ($d^{obs}$ frequency $< f_3$) inversion, U-net3 is applied to unflood the salt to its base, followed by a final FWI to fine-tune the model. Figure~\ref{fig:workflow} illustrates the case where $\mathbf{m_{init}}$ is a constant velocity; however, we will also examine the case where $\mathbf{m_{init}}$ is linear velocity $\mathbf{m(z)}$. 
\\
\\
The input data for training U-net1, U-net2, and U-net3 are 1D velocity profiles obtained by FWI using the different frequency bandwidths: first frequency bandwidth $\mathbf{m_{f_1}}$, second-frequency bandwidth $\mathbf{m_{f_2}}$, and the third frequency bandwidth $\mathbf{m_{f_3}}$, respectively. The target data for U-net1 and U-net2 (pre networks) are flooded true models ($m_{flooded}$). The target data for Unet-3 (the unflooding network) are the true model ($\mathbf{m_{true}}$). Since the networks also aim to estimate the pre/post salt velocity, we choose the mean squared error as the loss and minimize the networks as follows:  
\begin{equation}
    \mathbf{L}_{U_1} = \norm{(\mathbf{U1(m_{f_1})}-\mathbf{m_{flooded}}}^2_2,
    \label{eq:loss1}
\end{equation}
\begin{equation}
    \mathbf{L}_{U_2} = \norm{(\mathbf{U2(m_{f_2})}-\mathbf{m_{flooded}}}^2_2,
    \label{eq:loss2}
\end{equation}
\begin{equation}
    \mathbf{L}_{U_3} = \norm{(\mathbf{U3(m_{f_3})}-\mathbf{m_{true}}}^2_2,
    \label{eq:loss3}
\end{equation}
where $\textbf{U}$ stands for U-net. We summarize the input, output, and usage of the three networks in Table~\ref{table:1}. We minimize the losses using an Adam optimizer with a learning rate of 0.0001. We use a batch size of 32 and run the training for 50 epochs. We generate 10,000 training data in all our experiments and apply an 80-20 \% split to the data for the training and validation sets, respectively. We use the coefficient of determination, also known as the $\textbf{R}^2$ score, as an accuracy metric, which is given by: 
\begin{equation}
     \textbf{R}^2 = 1 - \frac{\sum_i^N (m_{true}-m_{pred})^2}{\sum_i^N (m_{true}+\bar{m}_{true})^2},
\end{equation}
where $m_{pred}$ indicates the predicted flooding/unflooding by the network and $\bar{m}_{true}$ represents the mean value for the true models. $\textbf{R}^2$ ranges between -1 and 1, with -1 indicating that the variables are anti-correlated and 1 representing a perfect correlation. 
\begin{table}[ht!]
\begin{center}
\begin{tabular}{ |p{.12\linewidth}|p{.14\linewidth}|p{.14\linewidth}|p{.14\linewidth}|p{.14\linewidth}|}
 \hline
 \textbf{Network} & \textbf{Frequency bandwidth} & \textbf{Input} & \textbf{Output} & \textbf{Usage}  \\ 
  \hline
U-net1 & $1^{st}$ bandwidth & FWI($\mathbf{m_{init}}$) & $\mathbf{m_{flooded1}}$ & Flooding\\
\hline 
U-net2 & $2^{nd}$ bandwidth & FWI($\mathbf{m_{flooded1}}$) & $\mathbf{m_{flooded2}}$ & Flooding \\
\hline 
U-net3 & $3^{rd}$ bandwidth & FWI($\mathbf{m_{flooded2}}$) & $\mathbf{m_{true}}$ & Unflooding \\
\hline 
\end{tabular}
\caption{Summary of the three networks. FWI($\mathbf{x}$) indicates an FWI result starting with $\mathbf{x}$ model. }
\label{table:1}
\end{center}
\end{table}
\subsection*{Training dataset}
Training neural networks requires a huge amount of data. This means that we need to apply FWI to many velocity models to generate data for our networks. Normally, this is impractical as the FWI algorithm is not cheap for a single model. Thus, we consider laterally invariant models, which we refer to as 1D models, to simplify the process and apply a fast FWI. In 1D models, FWI can be applied using only one shot where a full fold behavior is obtained by laterally stacking the gradient \citep{alali-unflooding,alkhalifah2018singleshot}. The use of 1D models is justified as the flooding and unflooding are 1D processes.
\\
\\
The training dataset should cover all the plausible scenarios for the networks to generalize to an unseen dataset. This is a challenge in most geoscience applications as the Earth is so diverse and have many geological structures. Luckily, another advantage of using 1D models is limiting the degrees of freedom that could have arised from the lateral heterogeneity. We create around 10,000 random velocity models to train the networks. We only consider a marine setup where the first layer correspond to water of velocity of 1.5 km/s. We randomize the models by considering the following variables: depth of water bottom, having a salt or not, depth of salt, thickness of salt, position of salts, number of layers, smoothness of layers. We assume a velocity of 4.5 km/s for salt layers. We guide the mean of the velocities to generally increase with depth. We also incorporate some prior knowledge about the field of study, such as the expected range of velocities and the general trend ( slope) of increase with depth, or any other information that can be extracted from an available well. Thus, we try to ensure that the true model falls within the distribution of models we train with as much as possible and at the resolution we seek \citep{ben2006analysisDomain}.
\\
\\
Once the velocity models are created, we simulate one shot for each model to perform FWI. For consistency, we use the same wavelet, offset, and minimum frequency as that of the observed seismic data. We investigate two cases for the FWI initial models. The first case is a constant starting model. We test this case on the synthetic BP 2004 salt model. The second case is an increasing linear model with depth, which is tested on a challenging real data example.

\section*{Synthetic Example: BP 2004 salt model}
\begin{figure}[!htb]
    \centering
    \subfigure[]{\includegraphics[width=0.3\textwidth]{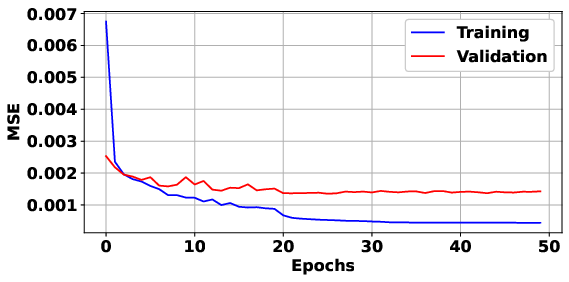}
    \label{fig:loss1}} 
    \subfigure[]{\includegraphics[width=0.3\textwidth]{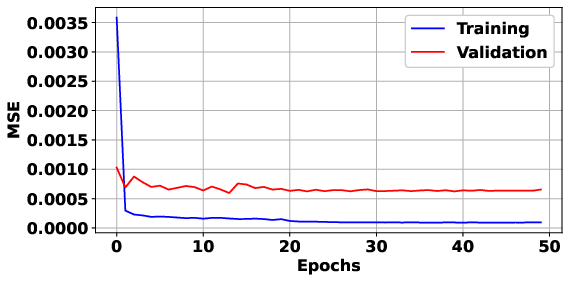}
    \label{fig:loss2}} 
    \subfigure[]{\includegraphics[width=0.3\textwidth]{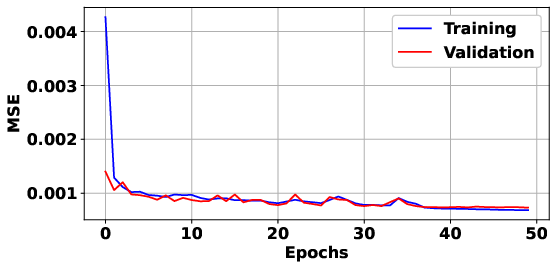}
    \label{fig:loss3}}     
    \subfigure[]{\includegraphics[width=0.3\textwidth]{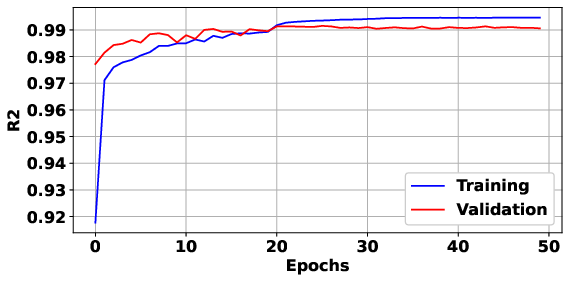}
    \label{fig:R2-1}} 
    \subfigure[]{\includegraphics[width=0.3\textwidth]{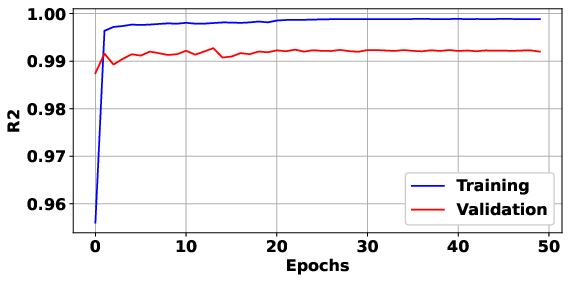}
    \label{fig:R2-2}} 
    \subfigure[]{\includegraphics[width=0.3\textwidth]{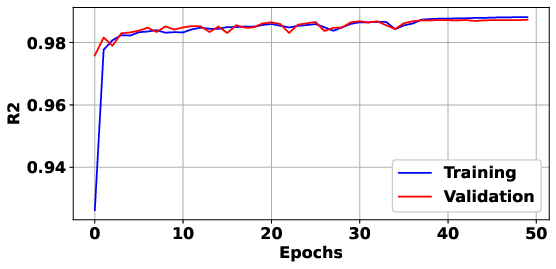}
    \label{fig:R2-3}}     
    \caption{Losses and R2 scores for the three networks. The column shows the loss and R2 score for U-net1, U-net2 and U-net3, respectively.}
    \label{fig:losses-R2}
\end{figure}

\begin{figure}[!htb]
    \centering
    \subfigure[]{\includegraphics[width=0.7\textwidth]{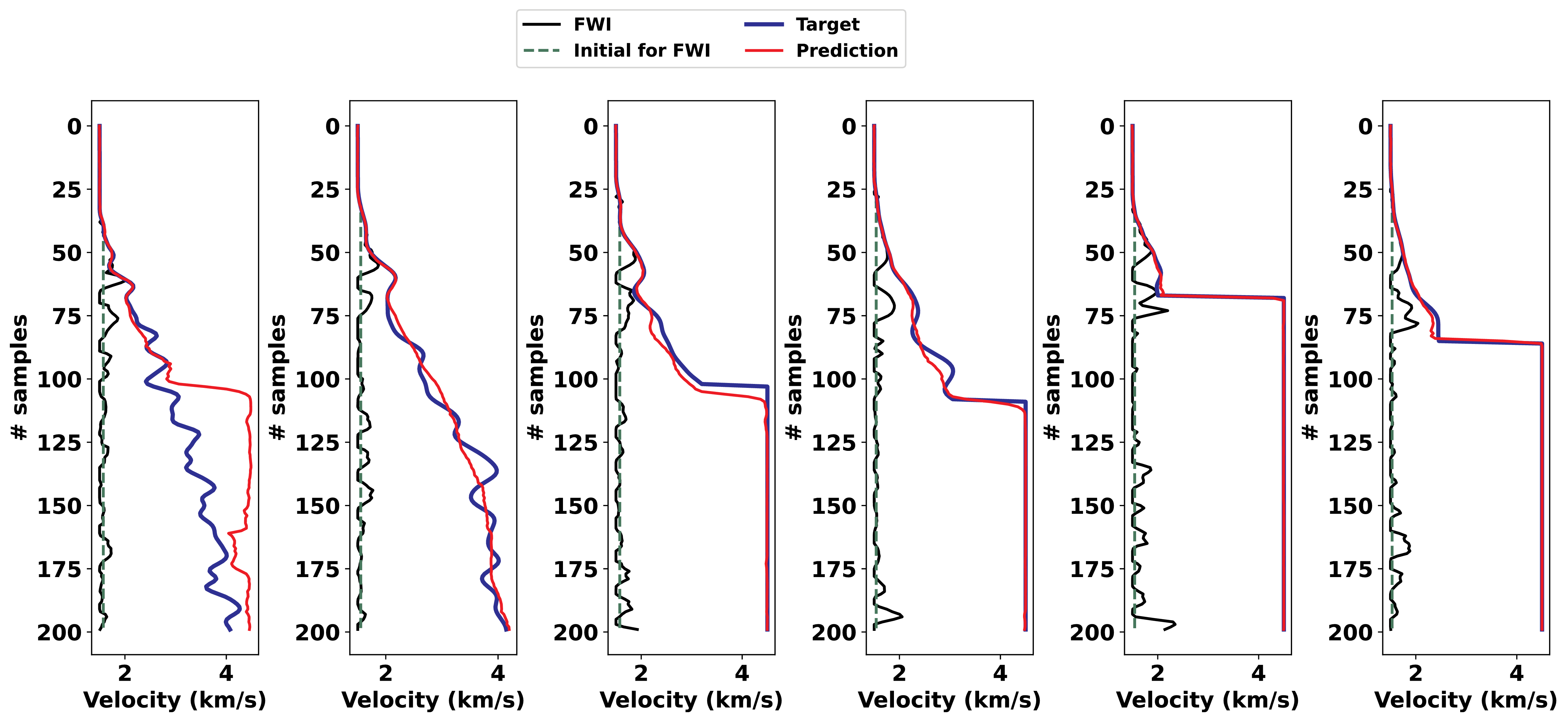}
    \label{fig:bp_val1}} 
    \subfigure[]{\includegraphics[width=0.7\textwidth]{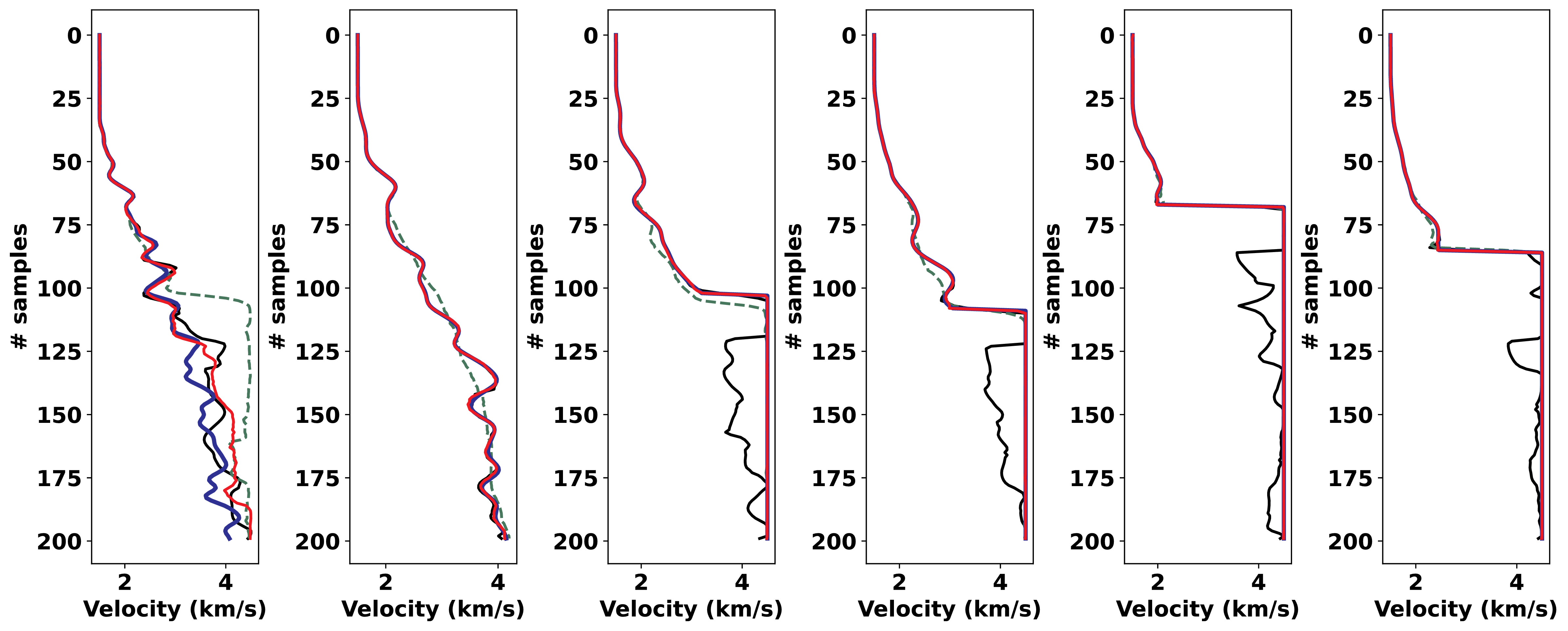}
    \label{fig:bp_val2}} 
    \subfigure[]{\includegraphics[width=0.7\textwidth]{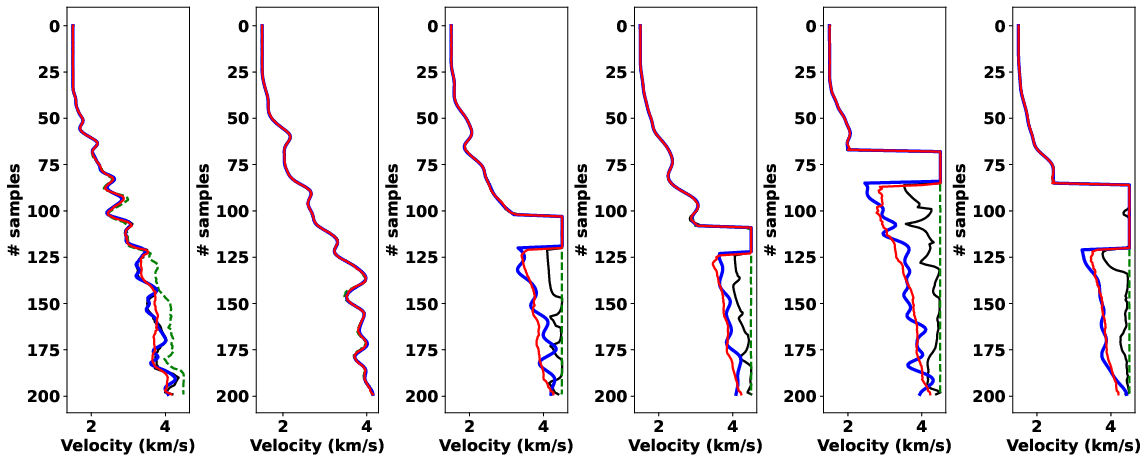}
    \label{fig:bp_val3}} 
    \caption{Samples from the validation sets for the three networks used to invert the BP 2004 salt model. a, b and c indicate the networks: U-net1, U-net2, and U-net3, respectively. The blue lines represent the true models. The black lines are the FWI results using the green dashed lines as the initial models. The red lines represent the predicted flooding/unflooding by the networks.}
    \label{fig:valbp}
\end{figure}

\begin{figure}[!htb]
    \centering
    \subfigure[]{\includegraphics[width=0.3\textwidth]{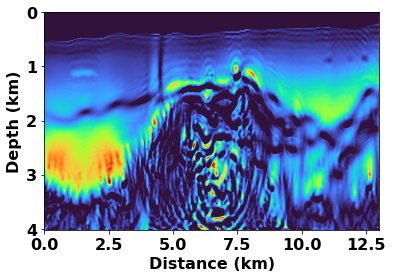}
    \label{fig:bpinv1}} 
    \subfigure[]{\includegraphics[width=0.3\textwidth]{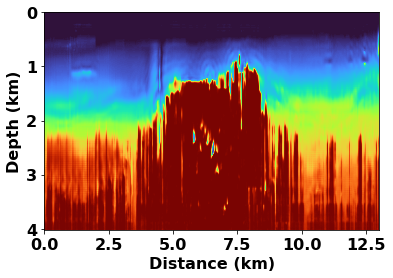}
    \label{fig:bpf1}} \\
    \subfigure[]{\includegraphics[width=0.3\textwidth]{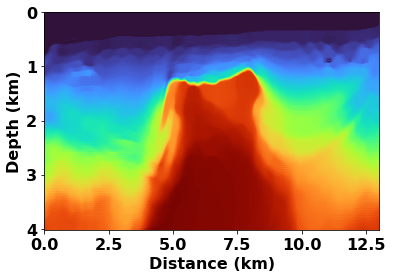}
    \label{fig:bpinv2}} 
    \subfigure[]{\includegraphics[width=0.3\textwidth]{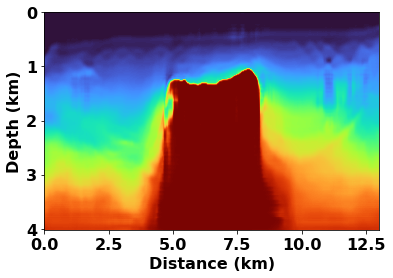}
    \label{fig:bpf2}} \\
    \subfigure[]{\includegraphics[width=0.3\textwidth]{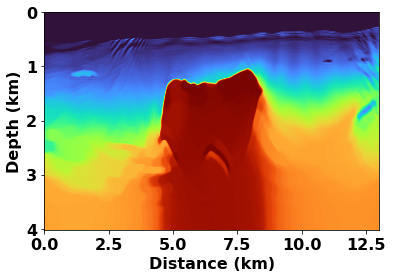}
    \label{fig:bpinv3}} 
    \subfigure[]{\includegraphics[width=0.3\textwidth]{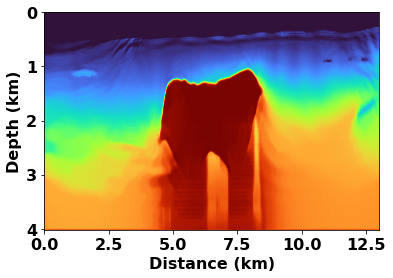}
    \label{fig:bpuf}} \\
    \subfigure[]{\includegraphics[width=0.3\textwidth]{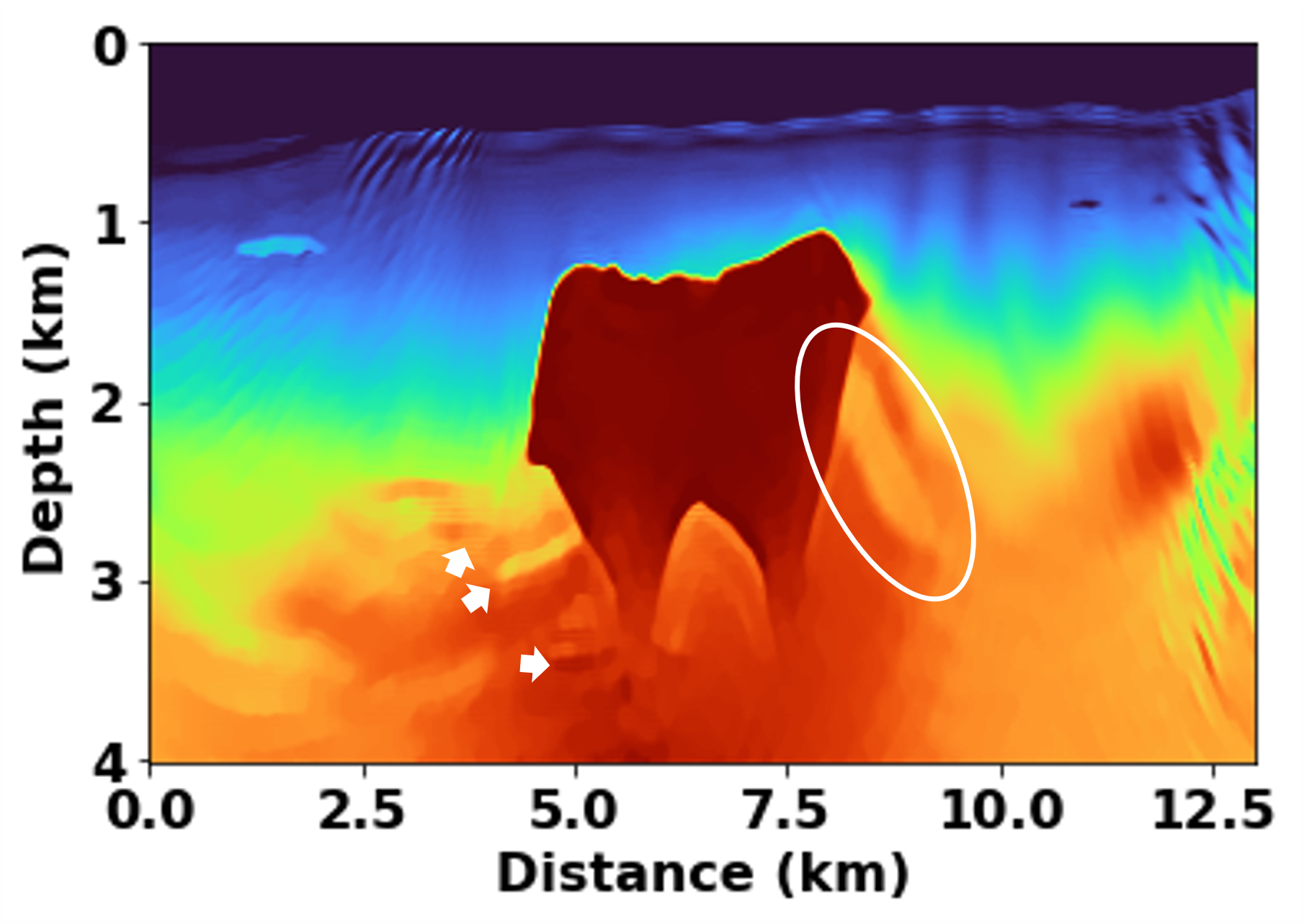}
    \label{fig:bpinvf}} 
        \subfigure[]{\includegraphics[width=0.35\textwidth]{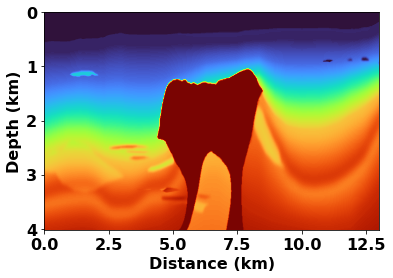}
    \label{fig:bptrue}} 
    \caption{BP inversion result using the proposed workflow. a) is the first FWI result starting from a constant velocity model. b) is the model after applying U-net1 for the first flooding. c) and d) are the subsequent FWI and the application of U-net2 for the second flooding, respectively. e) shows the following FWI used to unflood the salt and the result of applying Unet-3 for unflooding is shown in f. g) is the final FWI result starting from f. h) is the true model. The arrows point to areas where the inverted model matched the true one at depth.}
    \label{fig:bp}
\end{figure}

We test our workflow on the synthetic BP 2004 salt benchmark model. To reduce the computational cost, we crop the original model to only the central part and downsample the model to half its original size. We use a 5 Hz Ricker wavelet to simulate 144 shots placed at the surface and spaced 55 m apart. The maximum offset for the streamer is 6 km. We deploy 300 receivers at the surface with 40 m spacing. We apply a high-pass filter to remove frequencies below 3 Hz. In this experiment, we start the inversion from a constant velocity model of about 1.6 km/s, corresponding to that of the water bottom velocity. We perform a multi-scale FWI using low-pass filters with the following cut-off frequencies: 7, 10, and 15 Hz. 
%
%
\\
\\
Based on this survey and inversion parameters, we create our training data, as explained in the previous section. The loss and $\textbf{R}^2$ scores for the three networks are shown in Figure~\ref{fig:losses-R2}. All the losses converge relatively fast with high $R^2$ scores larger than 98\%. Figure~\ref{fig:bp_val1} shows U-net1 performance on six validation samples. The black lines show the inversion results using the first frequency scale starting from a constant velocity model shown in the green lines. Given the poor starting model and the lack of low frequencies, the inversion fails to recover the true model even when there is no salt, such as in the first two samples. Applying U-net1 improves the inversion dramatically for all the models and predicts initial flooding for the salts, as shown by the red lines. Comparing the predicted results with the true models (blue lines), we can observe that the network predicts the post-salt velocity fairly well. However, the predicted flooding is inaccurate in some cases, such as the false flooding in the first sample or the slightly deeper flooding in the third sample. This encourages us to perform another FWI and flooding step starting from the prediction results, which serve as improved initial models. Figure~\ref{fig:bp_val2} shows the result for the second flooding (red lines) by applying U-net2 on the second inversion scale (black lines). It is clear that the second flooding is more accurate than the previous one. In fact, most of the improvements occurred in the FWI stage, and the network helped sharpen the salts and flood them again. Notice how even the false flooding in the first sample was corrected, which is something we will exhibit in the 2D inversion  later in this section. Finally, we apply the third inversion scale and unflooding using U-net3 to these samples in Figure\ref{fig:bp_val3}. We can see that we achieved a good reconstruction for the salt models and the subsalt velocity in these samples. 
\\
\\
We show the first inversion result (frequencies < 7 Hz) for the BP salt model in Figure~\ref{fig:bpinv1}. The inverted model is far from the true model courtesy of starting from a constant velocity model, short offsets, and lack of low frequencies. We can see high energy around the top of the salt and little on the side. The result of applying U-net1 to the inverted model is shown in Figure\ref{fig:bpf1}. The network floods the salt with some inaccuracy, similar to what we observed in the five validation samples. We also see some false flooding on the sides of the salt. The predicted model exhibits vertical stripping due to the network trained on 1D. To mitigate the effects of the vertical strips, we slightly smooth the model laterally before running the next inversion scale. Going through the workflow and applying the subsequent inversion (frequencies < 10 Hz), we obtain the results shown in Figure\ref{fig:bpinv2}. We notice that all the false flooding are removed and the salt structure is more mature. Flooding the model again results in sharp and more accurate flooding (Figure~\ref{fig:bpf2}). Running the third inversion scale (frequency < 15 Hz) reveals evidence of the bottom of the salt (Figure~\ref{fig:bpinv3}) that is detected by U-net3 and used to unflood the salt (Figure~\ref{fig:bpuf}). The salt body is almost reconstructed at this stage, but we need a final FWI to fine-tune the model. The final inversion result is shown in Figure~\ref{fig:bpinvf}. Comparing the result with the true model given in Figure~\ref{fig:bptrue}, we can clearly see that we successfully build the salt body. We also recover some fine structures below the salt as indicated by the arrows and a low-velocity region indicated by an oval.    
\section*{Field example}
\begin{figure}[!htb]
    \centering
    \sbox{\measurebox}{%
    \begin{minipage}[b]{.3\textwidth}      \subfigure{\label{fig:shot}\includegraphics[width=0.9\textwidth,height=7cm]{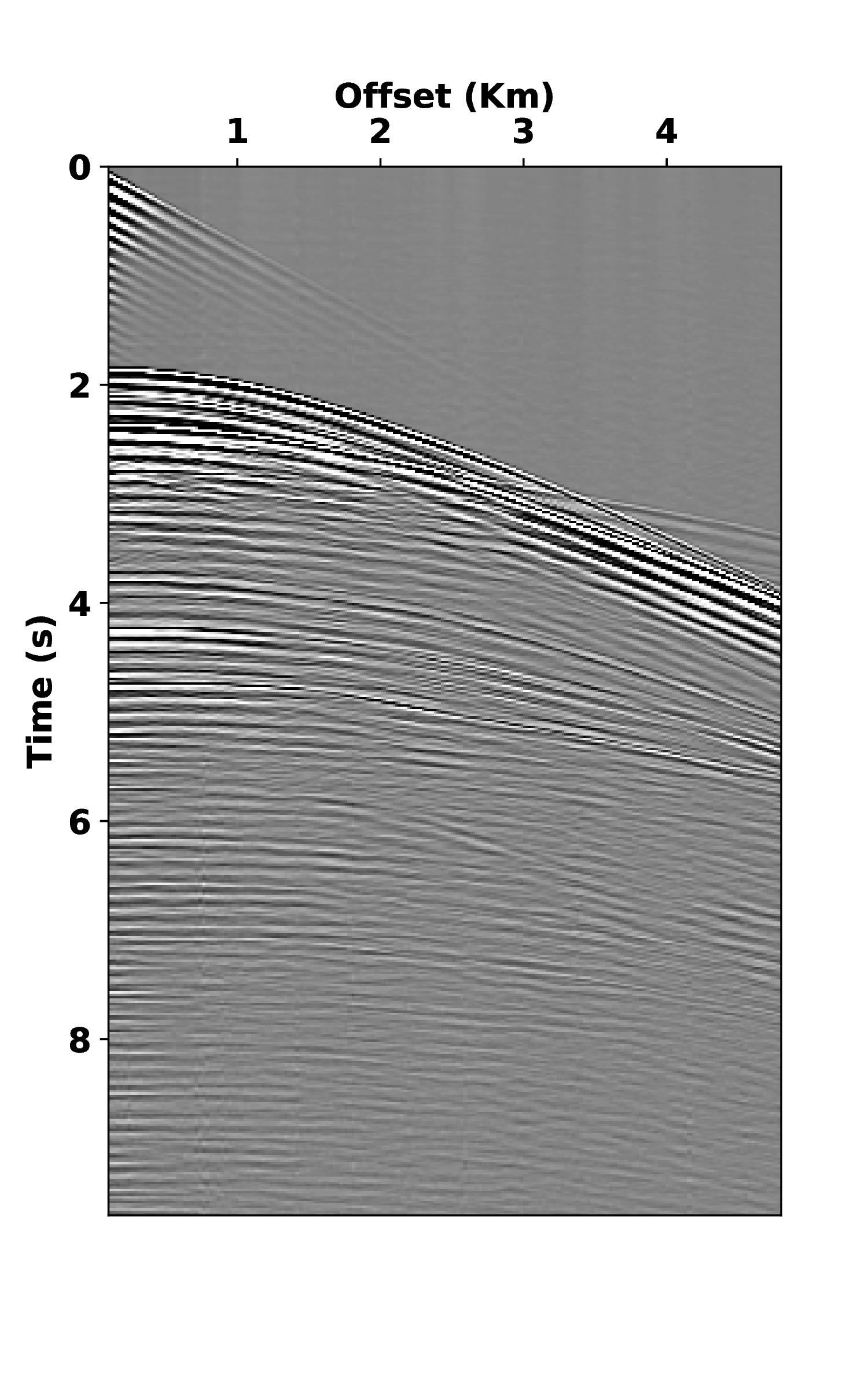}}
    \end{minipage}}
    \usebox{\measurebox}\qquad
    \begin{minipage}[b][\ht\measurebox][s]{.5\textwidth}
        \centering
        \vspace{25pt}        \subfigure{\label{fig:spectrum}\includegraphics[width=0.9\textwidth,height=2.5cm]{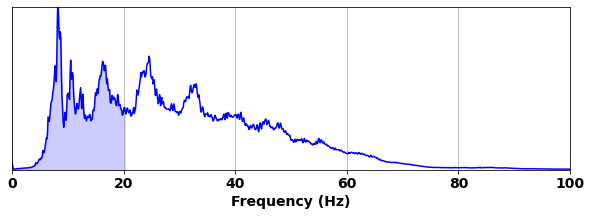}}
    \subfigure{\label{fig:spectrum_zoom}\includegraphics[width=0.9\textwidth,height=2.5cm]{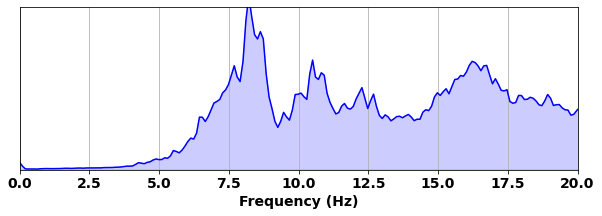}}
    \end{minipage}
\caption{A shot gather for the GOM data located at position x=10 km and its frequency spectrum. The bottom spectrum is a zoom in the shaded part. The plot shows a maximum offset of about 4.8 km and a minimum frequency of around 6 Hz.}
\label{fig:shot_spectrum}
\end{figure}

\begin{figure}[!htb]
    \centering
    \subfigure[]{\includegraphics[width=0.7\textwidth]{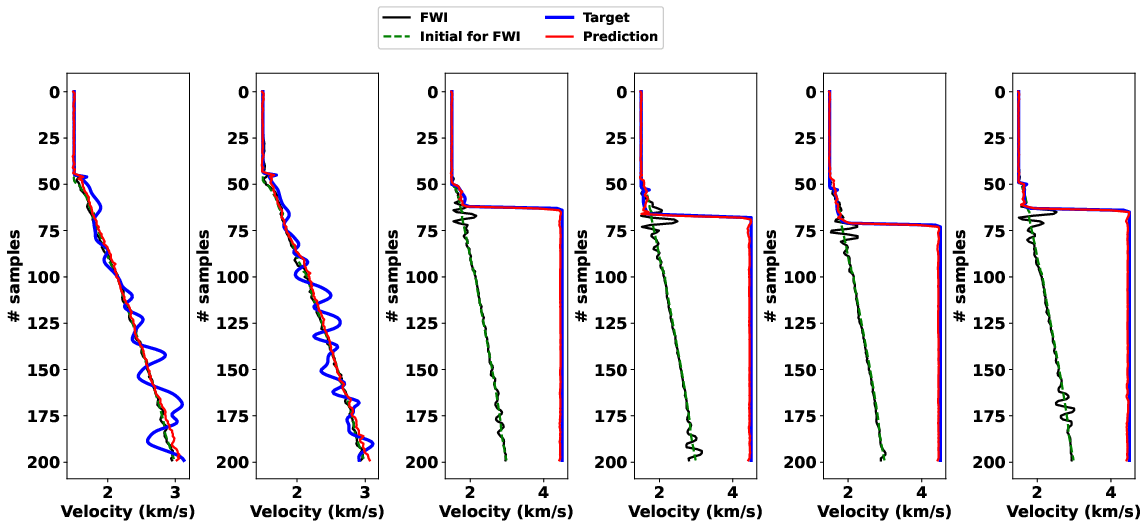}
    \label{fig:gom_val1}} 
    \subfigure[]{\includegraphics[width=0.7\textwidth]{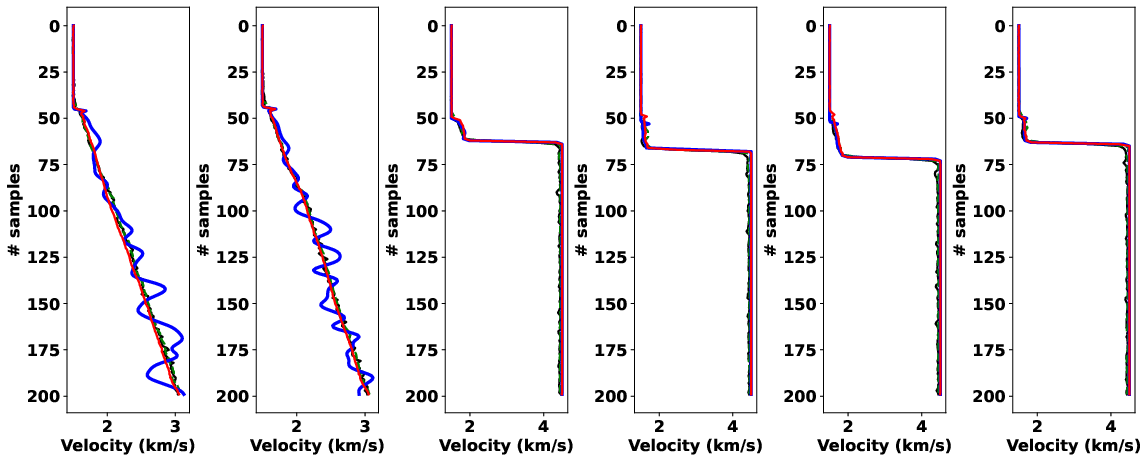}
    \label{fig:gom_val2}} 
    \subfigure[]{\includegraphics[width=0.7\textwidth]{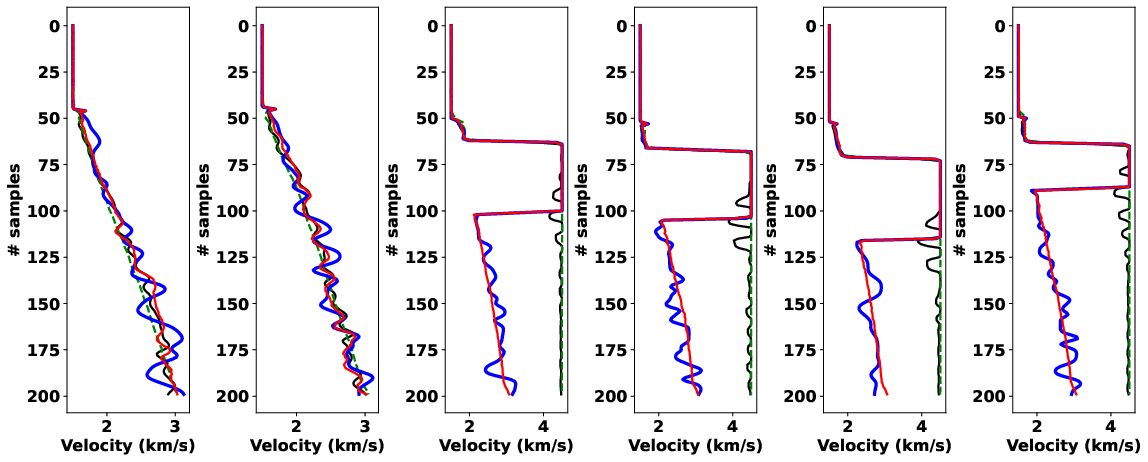}
    \label{fig:gom_val3}} 
    \caption{Samples from the validation sets for the three networks used to invert for the GOM salt model. a, b and c indicate the networks: U-net1, U-net2, and U-net3, respectively. The blue lines represent the true model. The black lines are the FWI results using the green dashed lines as the initial models. The red lines represent the predicted flooding/unflooding by the networks.}
    \label{fig:valgom}
\end{figure}

\begin{figure}[!htp]
	\centering
	\subfigure[]{\includegraphics[width=0.45\columnwidth]{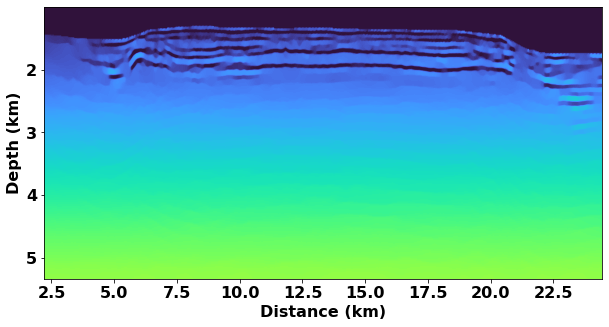}
		\label{fig:gominv1}}
	\subfigure[]{\includegraphics[width=0.45\columnwidth]{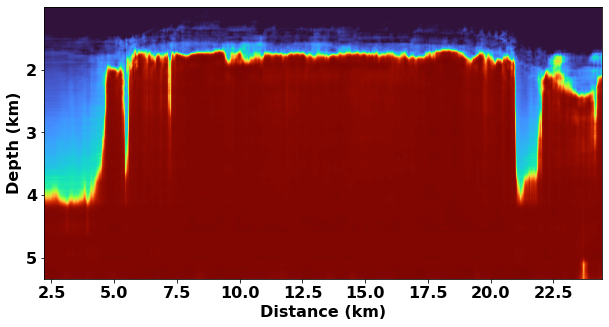}
		\label{fig:gomfl1}}
	\subfigure[]{\includegraphics[width=0.45\columnwidth]{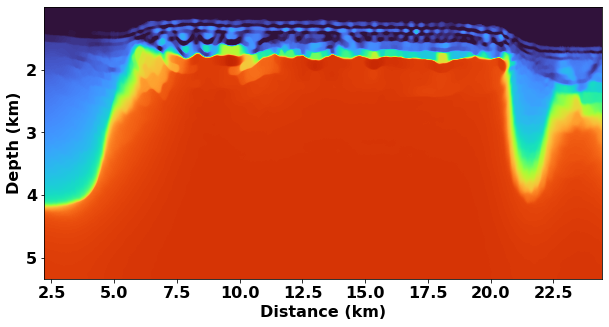}
		\label{fig:gominv2}}
	\subfigure[]{\includegraphics[width=0.45\columnwidth]{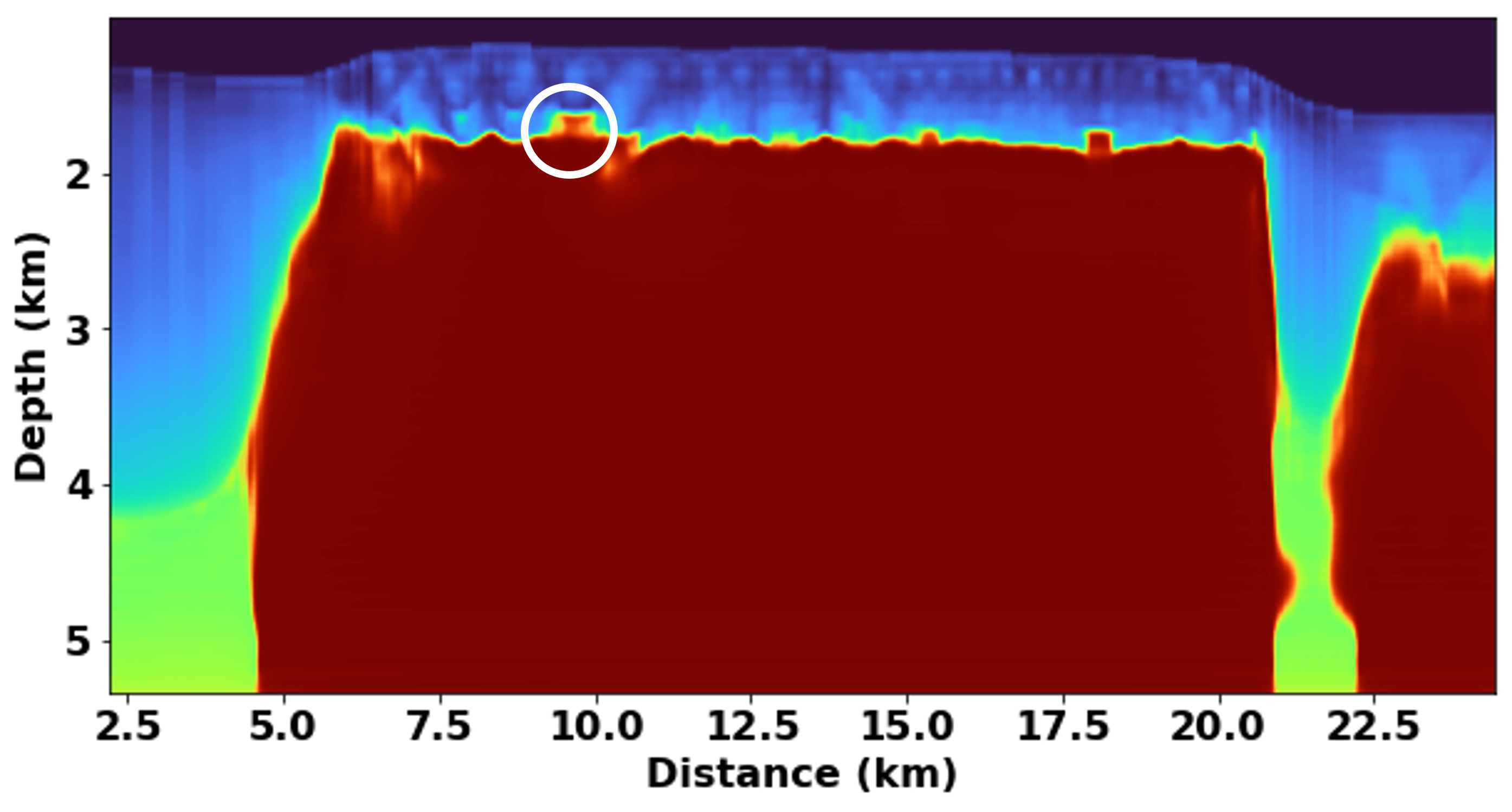}
		\label{fig:gomfl2}}
	\subfigure[]{\includegraphics[width=0.45\columnwidth]{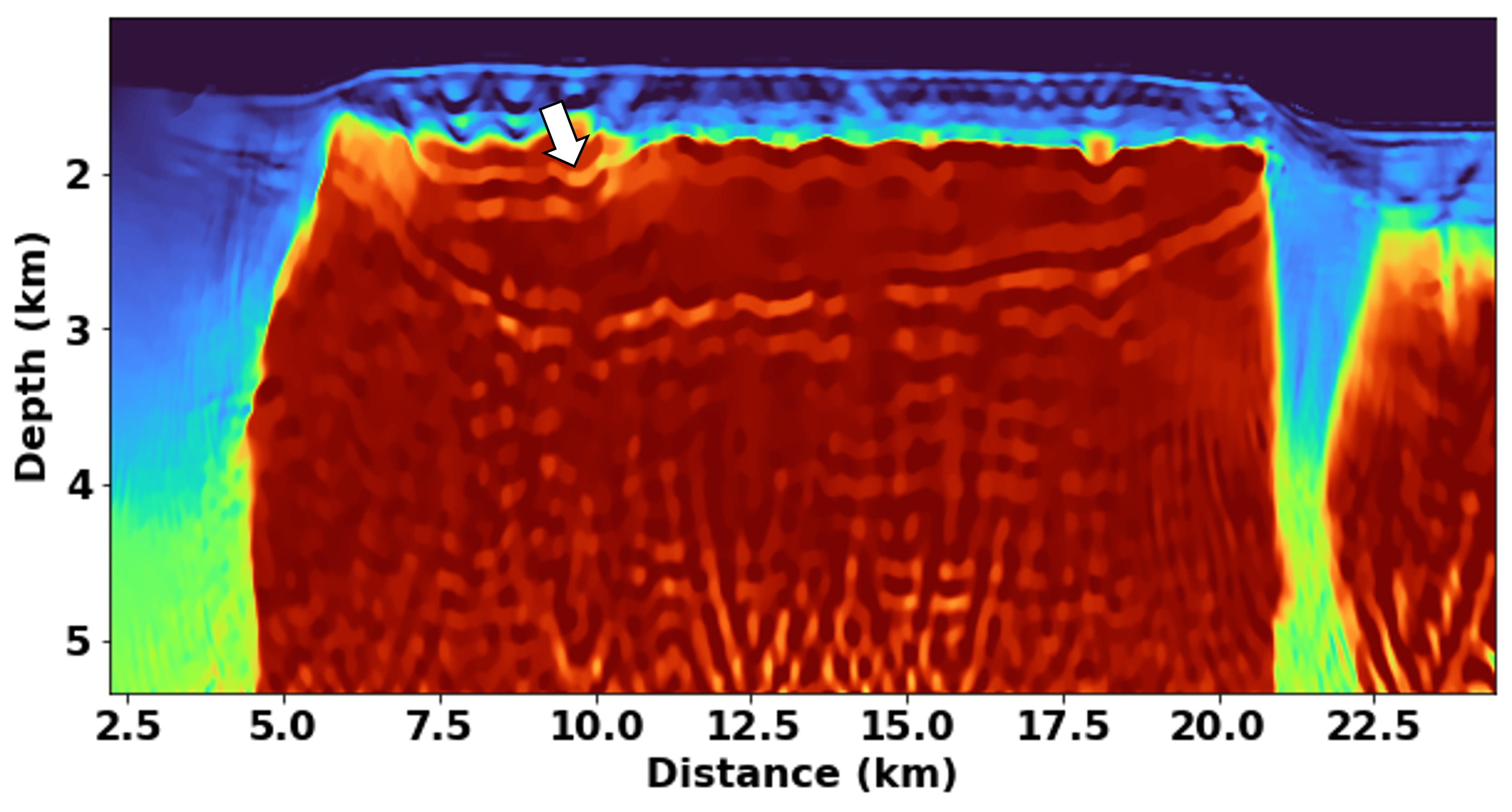}
		\label{fig:gominv3}}
	\subfigure[]{\includegraphics[width=0.45\columnwidth]{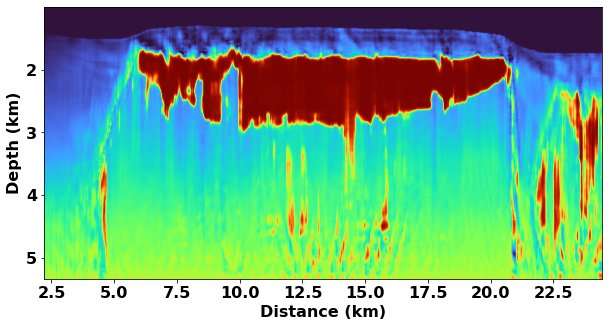}
		\label{fig:gomuf}}
		\subfigure[]{\includegraphics[width=0.45\columnwidth]{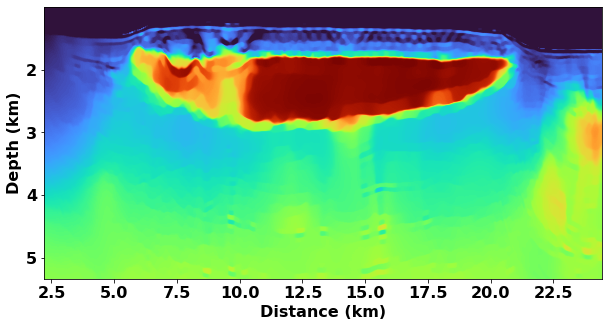}
		\label{fig:gomvf}}
	\caption{(a) is the first FWI with frequencies up to 7.5 Hz starting from a linear model $\mathbf{m(z)}$, (b) contains the predicted flooded model by U-net1, (c) is the FWI result starting from the flooded model with frequencies less than 10 Hz, (d) is the second predicted flooding by U-net2, (e) is the FWI result starting from (d) with frequencies less than 15 Hz, (e) is the predicted unflooded model by U-net3, and (g) is the final FWI result.}
	\label{fig:GOM_inversion}
\end{figure}

\begin{figure}[!htp]
	\centering
		\subfigure[]{\includegraphics[width=0.6\columnwidth]{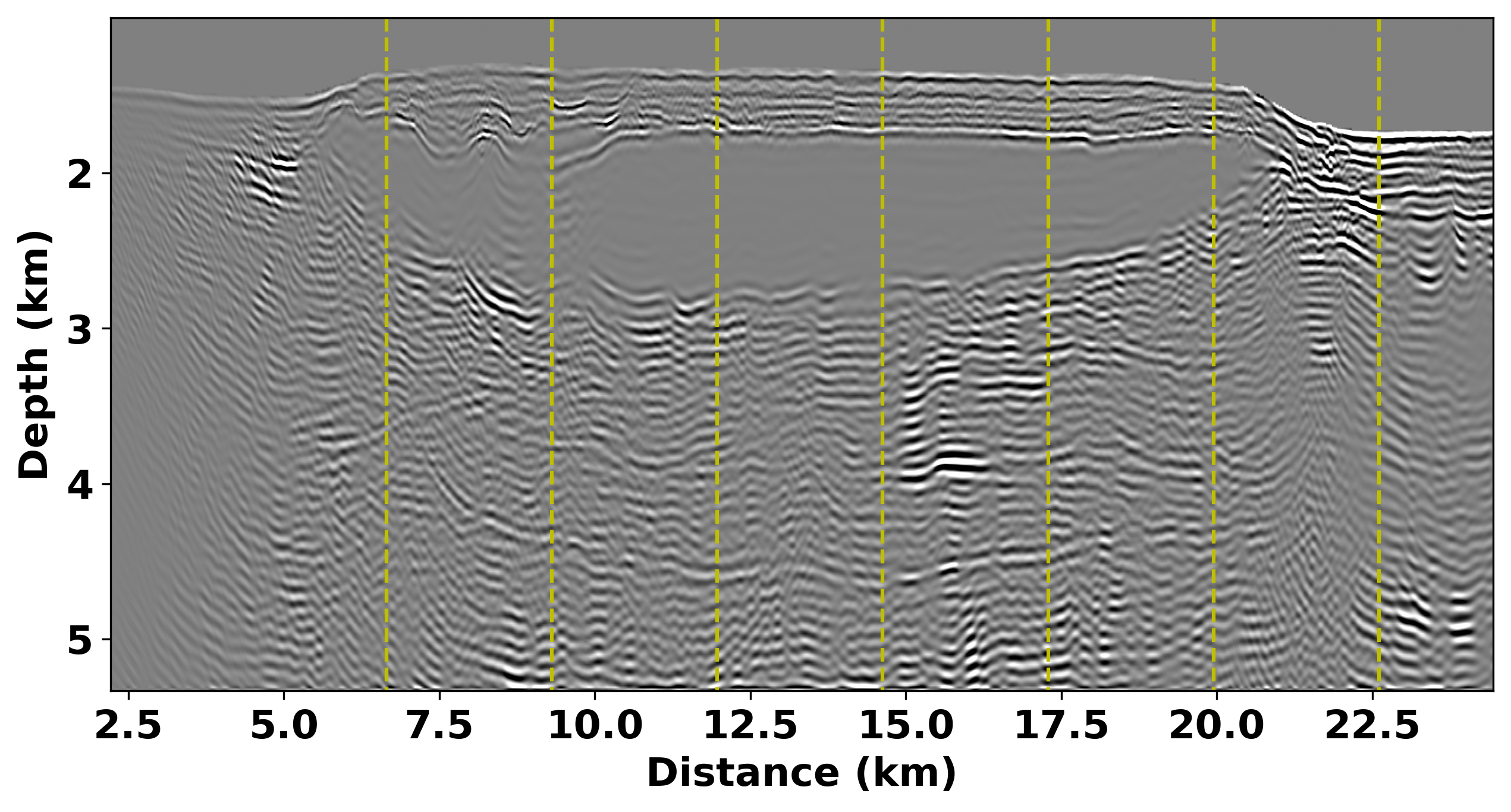}
		\label{fig:gomrtm}}
		\subfigure[]{\includegraphics[width=0.6\columnwidth]{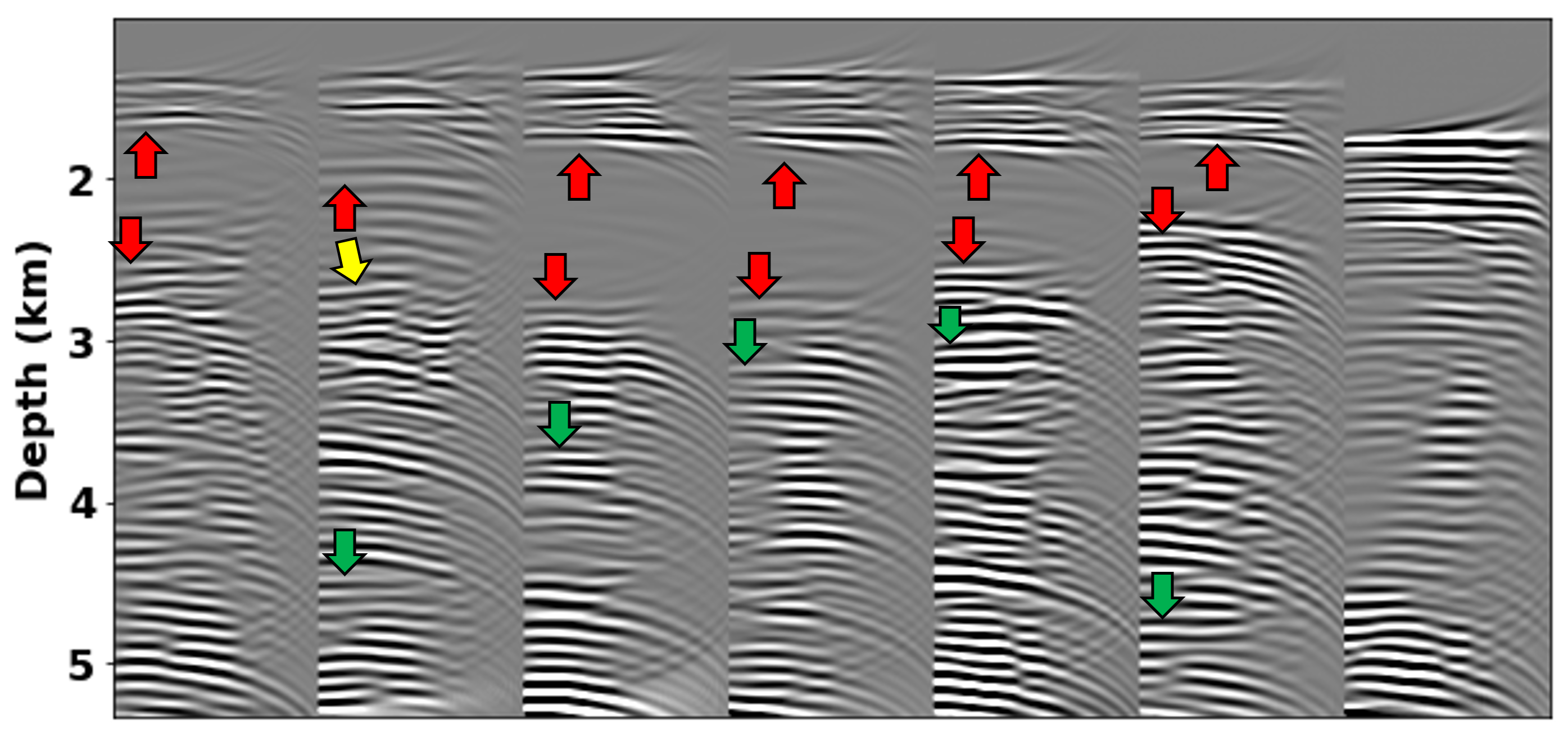}
		\label{fig:gomcig}}
            
	\caption{(a) RTM image using the final FWI velocity and (b) is eight common angles gathers panels taken from the positions indicated by the yellow dashed lines in (a).}
	\label{fig:GOM_imaging}
\end{figure}

We apply the method to a 2D dataset acquired in the GOM region, known as the Mississippi Canyon dataset. The data were acquired sometime in the nineties when exploration often involved high frequencies and short offsets. Most of the advanced acquisitions used in salt fields nowadays were not developed. We show a sample shot gather from the dataset in Figure~\ref{fig:shot_spectrum} and its corresponding spectrum, indicating a minimum usable frequency of about 6 Hz. In addition, the maximum available offset is 4.8 km. The data are known to have a shallow salt body in a deep water environment. The shallow salt results in severe surface multiples in the recorded data, which increase the probability of falling into local minima in the inversion \citep{guitton1999multiple}. 
\\
\\
\\
We create a 1D training dataset using the same survey parameters used in this field. This time, we start the inversion using a linearly increasing velocity with depth \textbf{m(z)} at a gradient of 0.005 $s^{-1}$ starting from the water bottom. Similar to the synthetic model, we show six validation samples for all three networks in Figure~\ref{fig:valgom}. The predictions for U-net1 (first row) and U-net2 (second row) are not as different as we observed in the synthetic example. This is due to starting from \textbf{m(z)}, which is closer to the true random models (labels). The unflooding prediction is plotted in the last row, which shows a fairly good unflooding for the salt base on the validation data. Thus, the trained U-net models are generally accurate and ready to be used on the field data.   
\\
\\
We start the inversion for the GOM model using a linear velocity like that used in the training. We use multi-scale frequencies with the following high cut-off: 7.5, 10, and 15 Hz. The first inversion bandwidth is shown in Figure~\ref{fig:gominv1}. We observe high-energy events, which probably represent the top of the salt. After applying the first flooding network, we obtain Figure~\ref{fig:gomfl1}. Apparently, the network identifies the high-energy parts and floods from there. The first flooding identifies two salt structures one in the center and the other on the right side. We then proceed with the next frequency bandwidth and apply FWI followed by another flooding and plot the result in Figures~\ref{fig:gominv2} and \ref{fig:gomfl2}, respectively. The salt top is fine-tuned with this step and it looks more continuous. The next FWI using the third frequency range 
 is shown in Figure~\ref{fig:gominv3}, which clearly produces energy within the salt that we interpret as the bottom of the salt. The unflooding nicely detects this base and unfloods the salt velocity. However, at 10 km lateral location , there is a discontinuity in the shape of the salt. This is because the network picks the first low-velocity event indicated by the white arrow in Figure~\ref{fig:gominv3}. Tracking this point back to Figure~\ref{fig:gomfl2}, we also realize that this low velocity corresponds to a vertical pull-up indicated by the circle. The right side salt is mostly removed, probably because it does not show a clear base signature. It is hard to say for sure whether there is a side salt or not, but the algorithm initially predicts it due to the strong event representing the top of the salt in Figure~\ref{fig:gominv1}. Finally, we apply a last inversion on the unflooded model and obtain Figure~\ref{fig:bpinvf}. The inversion reveals a pillow-like salt shape. The salt exhibits low amplitudes in the left part, especially at about 10 km  lateral position. We suspect the complexity of the salt's top on the left part produces multi-scatter reflections that project on the salt reducing its amplitude. 
 \\
 \\
 To assess the accuracy of the inversion, we perform RTM on the final FWI model and obtain angle gathers. Flattened gathers indicate accurate velocities. The RTM image in Figure~\ref{fig:gomrtm} shows good energy focusing on the salt boundaries, especially on the right side of the salt. On the left side, where the salt top sinks deeper, the scattering of energy on the sides results in an unfocusing behavior, which we observe at about 7.5 and 10 km lateral location. The angle gathers in Figure~\ref{fig:gomcig} are computed along the yellow dashed lines in Figure~\ref{fig:gomrtm}. Most of the reflection energy corresponding to salt tops and bottoms are flattened as indicated by the red arrows, which verify the credibility of the inversion result. The third panel crosses the 9.3 km position within the sunken region. Examining this particular panel, we see good focusing at the salt's top as indicated by the red arrow, but the base depicts a lower velocity by showing an upward concavity indicated by the yellow arrow. The deeper parts (pre-salt) angle gathers indicate reflections experiencing higher velocities and that is mostly due to the surface and sea bottom multiple nature of such energy. For imaging, surface-related multiple filters can remove some of those. Otherwise, if we look closely we can identify a lot of flattend, slightly weak, subsalt (pre-salt) energy, which indicates the accuracy of the velocity model above it. We point to these flattened reflection energy by the green arrows. The last angle gather panel is taken at the area where the algorithm initially predicts salt and then removes it. The flatten events on all the panel shows that indeed there is no salt in that region. In general, we consider the inversion to be fairly good given the acquisition limitations: absence of low frequencies, short offsets, and a poor starting model.           
\section*{Discussion}
Our workflow tackles the limitation of the conventional top-down approach in building a salt velocity model. Those limitations can be summarized by: the manual picking of salt boundaries, and the cost of high-frequency advanced imaging. We achieved full automation of the process by utilizing U-nets. Unlike most of the studies in automating salt body building by picking the salt boundaries from images, we build the salt within the FWI process starting from low frequencies, which require coarser grids than applying high-frequency imaging. As a result, the cost of the proposed method is generally the cost of applying a conventional multi-scale FWI since the network application is proposed in the transition between frequency scales in FWI.  
\\
\\
\\
One crucial factor in our implementation is including TV regularization. Without TV, FWI fails to construct the salt as it will lack the model regularization push to do so. This phenomenon is exacerbated by the offset and frequency limitations in our applications. We implement TV regularization and control its amount by $\lambda$ in equation~\ref{eq:fwi}. In the early steps, where the salt is not mature, we chose $\lambda$ to be relatively large to keep the salt generally intact and slightly modify its structure. Then as we proceed in the workflow, especially in the unflooding step, we reduce $\lambda$ to allow for more details in the model, such as the salt base energy. The strategy of weakening the TV in FWI is not new and has been used in many studies \citep{kalita2019regularized,esser2018total}.    
\\
\\
Examining the workflow results step by step, we found that the approach includes somewhat of a self-correction mechanism. This is apparent in the synthetic example in the first validation samples in Figure~\ref{fig:valbp}. The first network prediction shows a false salt, which then is removed by the next FWI. Similarly, in the synthetic BP model, we see false flooding on the side of the salt (Figure~\ref{fig:bpf1}), which is subsequently removed by the following steps. In the GOM data, we observe a salt on the right side of the model that is also removed in the unflooding step in Figure~\ref{fig:gomuf}. The last angle gather panel in Figure~\ref{fig:gomcig} crosses the region of the right side salt, and it shows nice flattening. This means that the velocity is kinematically accurate in that region, and hence the initially predicted salt is indeed a false prediction.   
\\
\\
Nevertheless, if at the final step (the unflooding) the network exhibits a false prediction, which is seemingly the case for the GOM inversion at around 10 km lateral location, it might be useful to go over one additional round of flooding-unflooding and iterate between these steps until the salt model converges. This will require training an additional network to flood models that contain mature salts. We want to emphasize that in our application, we used vintage data that lack the information to invert for the salt due to its limited offset and frequencies. Thus, when using data corresponding to an advanced acquisition, FWI will most probably correct such false predictions much better.   
\\
\\
Courtesy of applying 1D flooding/unflooding, we can observe vertical stripping in our 2D predictions. The next step for this work could be to accommodate the 2D nature of our model. This could be done at the network level or as post-processing step. In our implementation, we apply a slight lateral smoothing to the network's prediction before running FWI, which we believe is sufficient enough to obtain reasonable results in our applications.
\section*{Conclusion}
We proposed a fully automated workflow to build salt bodies using FWI. We applied two consecutive 1D flooding and one unflooding using 1D U-net networks. Specifically, we train three networks: U-net1 for flooding the salt and improving the post-salt model, U-net2 is also for flooding to further correct the top of the salt, and U-net3 to unflood the salt to its base. The networks are implemented between frequency bandwidths in a multi-scale FWI approach. Owing to the ability of our framework to inject Salt body information into the model using neural networks, we test our approach on two (synthetic and field) challenging data of limited offset and bandwidth, and limited initial velocity information. We successfully manage to recover the salt information in both examples despite the data limitations.

\bibliographystyle{chicago}  
\bibliography{ref}

\end{document}